\documentclass[eqsecnum,aps,showpacs,amsmath,nofootinbib,superscriptaddress,preprintnumbers]{revtex4}

\usepackage{epsfig, color,graphicx}
\usepackage{amsmath}
\usepackage{amssymb}
\newcommand{\nn}{\nonumber \\}
\newcommand{\bea}{\begin{eqnarray}}
\newcommand{\ena}{\end{eqnarray}}
\newcommand{\vs}[1]{\vspace{#1 mm}}
\newcommand{\hs}[1]{\hspace{#1 mm}}
\renewcommand{\a}{\alpha}

\renewcommand{\c}{\gamma}
\renewcommand{\d}{\delta}
\newcommand{\e}{\epsilon}
\newcommand{\s}{\sigma}

\newcommand{\la}{\lambda}
\newcommand{\pa}{\partial}
\newcommand{\p}[1]{(\ref{#1})}

\newcommand{\tr}{\tilde{r}}

\newcommand{\squaret}{\kern1pt\vbox{\hrule height 0.9pt\hbox{\vrule width
0.9pt\hskip 2pt\vbox{\vskip 5.5pt}\hskip 3pt\vrule width 0.3pt}\hrule height
0.3pt}\kern1pt}
\begin{document}

\preprint{KU-TP 059}
\title{Asymptotically AdS Charged Black Holes in String Theory with Gauss-Bonnet Correction
in Various Dimensions
}
\author{Nobuyoshi Ohta}\email{ohtan@phys.kindai.ac.jp}
\affiliation{Department of Physics, Kinki University, Higashi-Osaka,
Osaka 577-8502, Japan}

\author{Takashi Torii}\email{torii@ge.oit.ac.jp}
\affiliation{Department of General Education, Osaka Institute of Technology,
Asahi-ku, Osaka 535-8585, Japan}

\date{\today}

\begin{abstract}
We study charged black hole solutions in Einstein-Maxwell-Gauss-Bonnet theory with the dilaton
field which is the low-energy effective theory of the heterotic string.
The spacetime is $D$-dimensional and assumed to be static and plane symmetric with
the $(D-2)$-dimensional constant curvature space and asymptotically anti-de Sitter.
By imposing the boundary conditions of the existence of the regular black hole horizon and
proper behavior at infinity where the Breitenlohner-Freedman bound should be satisfied,
we construct black hole solutions numerically.
We give the relations among the physical quantities of the black holes such as
the horizon radius, the mass, the temperature, and so on.
The properties of the black holes do not depend on the dimensions qualitatively,
which is different from the spherically symmetric and asymptotically flat case.
There is nonzero lower limit for the radius of the event horizon below which
no solution exists.
The temperature of the black hole becomes smaller as the horizon radius is smaller
but remains nonzero when the lower limit is attained.

\end{abstract}



\maketitle

\section{Introduction}
\label{intro}

It is expected that quantum gravity plays the significant role
in studying physics at strong gravity such as a black hole singularity.
Since the ten-dimensional superstring theories or eleven-dimensional M-theory
are the leading candidates for the quantum theory of gravity, it is interesting
to use the low-energy effective theories of string/M theories to explore the problem.
The effective theories are the supergravities which typically involve
not only the metric but also the dilaton field (as well as several gauge fields).
It is known that there are additional correction terms of higher orders in the curvature
to the lowest effective supergravity action,
the Gauss-Bonnet (GB) term coupled to the dilaton field in heterotic string~\cite{MT}.

When the dilaton field is set to a constant, the total action consists of the cosmological
constant, Einstein-Hilbert and GB terms, which are the first three terms in the Lovelock
theory~\cite{Lovelock,Zumino}. There have been many works on the black hole
solutions in the Lovelock theory~\cite{GG,KMR,TYM,CNO}.
Solutions in $D=4$ with GB term have been discussed in \cite{TYM}.
Boulware and Deser~\cite{BD} discovered static, spherically symmetric black hole
solutions of such models in more than four dimensions. In the system with a negative
cosmological constant, black holes can have horizons with nonspherical topology such as
torus, hyperboloid, and other compactified submanifolds. These
solutions were originally found in general relativity and are
called topological black holes~\cite{bro84}. Topological black
hole solutions were studied also in the Einstein-GB theory~\cite{Cai}.
However, the dilaton is not considered in most of the works.
It is one of the important ingredients in string theory, so it is
significant to study how its presence modifies the solutions.
Unfortunately, it is not possible to find analytic solutions when the dilaton is present
and is coupled to the GB term. We have to study this by numerical analysis.
For related solutions,
see \cite{TM,BGO,MOS1,CGOO,MOS2,CGK,MOW}.

In a series of papers~\cite{GOT1,GOT2,OT3,OT4,OT5}, we have studied neutral black hole
solutions in the low-energy effective heterotic string, namely the Einstein theory
plus the GB term coupled to the dilaton field~\cite{MT}.
We have studied these solutions for asymptotically flat, anti-de Sitter (AdS)
and de Sitter cases with suitable cosmological constants,
and exhibited their properties including mass, temperature, entropy and their relations.
These neutral solutions are useful for studying modifications of the black hole solutions
by higher order effects induced by string theories.
In particular, we have studied how the singularity may be modified and showed that
they may be significantly modified. Also those solutions in AdS spacetime have been useful
to study the corrections to the ratio of shear viscosity to entropy density~\cite{vis,CNOS}.

Given this situation, it is natural to extend the black holes to charged ones.
In our more recent paper~\cite{OT6}, we have studied asymptotically flat charged black hole
solutions in these theories.
In fact, in the study of superconductors and superfluidity using the AdS/CFT correspondence,
charged black hole solutions play important roles~\cite{Gubser}.
Hence it is significant to extend our above asymptotically flat solutions
to charged AdS ones. We should also consider the inclusion of the dilaton field.
One of the purposes of this paper is to study asymptotically AdS charged black hole
solutions with higher order corrections as well as the dilaton field.

There is another reason for our study of the solutions in such a system.
In our above study of the neutral black holes~\cite{GOT1,GOT2,OT3,OT4,OT5},
we did not consider the higher derivative term of the dilaton field.
It is known, however, that such terms are also present in the effective theory~\cite{MT}.
Hence we should incorporate this term as well and study how this modifies the solutions.
We have studied this problem in the paper for asymptotically flat solutions~\cite{OT6},
and found that there is not much qualitative difference.
We will show that this is also true for the asymptotically AdS solutions.
Thus the results in our earlier papers should be useful.

This paper is organized as follows. In Sec.~\ref{sec2}, we first present the action
and give basic equations to solve for the system of Einstein-Maxwell-Gauss-Bonnet term (EMGB)
coupled to the dilaton field with a cosmological constant.
We then discuss symmetries which can be used to obtain solutions with different charge, mass
and cosmological constant, given a solution for a charge, mass and cosmological constant.
Boundary conditions and asymptotic behaviors are explored and parameter regions
for the existence of the desired black hole solutions are determined.
We show that the extreme black hole solutions with degenerate horizon can exist only
when parameters take special values.
In Sec.~\ref{sec3}, we choose the parameters for which we construct the black hole
solutions, and discuss how the neutral solutions are modified in the presence of
higher derivative terms of dilaton.
Then the results for charged black holes are given and discussed for $D=4,5,6$ and 10.
Section~\ref{CD} is devoted to conclusions and discussions.

\section{Dilatonic Einstein-Maxwell-Gauss-Bonnet theory}
\label{sec2}

\subsection{Action and basic equations}

We consider the following low-energy effective action for the
heterotic string theory in one scheme~\cite{MT}:
\bea
S &=& \frac{1}{2\kappa_D^2}\int d^{D}x \sqrt{-g} \Bigg[R -\frac12 (\pa \phi)^2
-\frac{1}{4}e^{-\c\phi} F^2 + \a_2 e^{-\c\phi} \Big\{ R^2_{\rm GB}
+ \frac{3}{16} (\pa\phi)^4 \Big\} -\Lambda e^{\la\phi} \Bigg],
\label{action1}
\ena
where $\kappa_{D}^2$ is a $D$-dimensional gravitational constant,
$\phi$ a dilaton field, $F$ a gauge field strength, $\c=1/2$, $\alpha_2=\a'/8$ is a numerical
coefficient given in terms of the Regge slope parameter $\a'$, $\la$ is a constant parameter, and
$R^2_{\rm GB} = R_{\mu\nu\rho\sigma} R^{\mu\nu\rho\sigma}- 4 R_{\mu\nu} R^{\mu\nu} + R^2$
is the GB term. The field redefinition ambiguity~\cite{MT,MOW} is applied
to put the action into the above form. The three-form field $H_{\mu\nu\rho}$ is set
to zero. This is allowed since it is a solution of the field equations.
We investigated the neutral black hole solutions in the similar system without Maxwell field
in Refs.~\cite{GOT1,GOT2,OT3,OT4,OT5}. However, we did not take into account
the higher order term of the dilaton field $(\partial \phi)^4$.
Here we briefly discuss the modification due to this term as well as the charged solutions.

To construct black hole solutions,
let us consider the metric and field strength:
\bea
ds_D^2 = - B(r) e^{-2\d(r)} dt^2 + B(r)^{-1} dr^2 + r^2 h_{ij}dx^i dx^j,~~
F_{0r}=\frac{df(r)}{dr},
\ena
where $h_{ij}dx^i dx^j$ represents the line element of a $(D-2)$-dimensional
constant curvature space with zero curvature and volume $\Sigma_0$.

The field equations following from the action~\p{action1} are
\bea
&& \bigl(B\tr^{D-3}\bigr)' \frac{D-2}{\tr^{D-4}}h +\frac12 B \tr^2 {\phi'}^2
 + (D-1)_4\,e^{-\c\phi}\frac{B^2}{\tr^2} + 4(D-2)_3\, \c e^{-\c\phi}B^2(\phi''-\c {\phi'}^2) \nn
&& \hs{5} - 2(D-1)_3\,\c e^{-\c\phi}\phi'\frac{B^2}{\tr}
+\frac{\tr^2}{2} e^{2\d-\c\phi}f'^2 - \frac{3}{16} B^2 \tr^2 e^{-\c\phi}\phi'^4
+\tr^2 \tilde{\Lambda}e^{\lambda\phi} = 0\,,~~
\label{fe1}
\ena
\bea
\delta'(D-2)\tr h + \frac12 \tr^2 {\phi'}^2
+ 2(D-2)_3\, \c e^{-\c\phi} B(\phi''-\c {\phi'}^2)
-\frac{3}{8} \tr^2 B \phi'^4 e^{-\c\phi}=0 \,,
\label{fe2}
\ena
\bea
(e^{-\d} \tr^{D-2} B \phi')' &=& \c (D-2)_3 e^{-\c\phi-\d} \tr^{D-4}
\Big[ (D-4)_5 \frac{B^2}{\tr^2} + 2(B'-2\d' B)B' +4B^2 U(r)
+4\frac{D-4}{\tr}(B'-\d'B)B \Big] \nn
&& +\c \tr^{D-2} e^{-\c\phi} \Big(\frac{1}{2}e^{\d}f'^2
+ \frac{3}{16} e^{-\d} B^2\phi'^4\Big) \nn
&& + \frac{3}{4} \tr^{D-2} B\phi'^2 e^{-\d-\c\phi}\Big\{3B\phi''
+\Big[2B'-\Big(\d'+\c\phi'
-\frac{D-2}{r}\Big)B\Big]\phi' \Big\}
+ \tr^{D-2}\lambda \tilde{\Lambda}e^{-\delta+\lambda\phi},
\label{fe3}
\ena
\bea
(f'e^{\d-\c\phi}\tr^{D-2})'=0,
\label{fe4}
\ena
where we have defined the dimensionless variables: $\tr = r/\sqrt{\a_2}$,
$\tilde \Lambda = \a_2 \Lambda$, and the primes in the field equations
denote the derivatives with respect to $\tr$. Namely the length is measured
in the unit of $\sqrt{\a_2}$.
In what follows, we omit tilde on the variables for simplicity.
We have also defined
\bea
(D-m)_n &=& (D-m)(D-m-1)(D-m-2)\cdots(D-n), \nn
\label{h-def}
h &=& 1+2(D-3) e^{-\c\phi} \Big[ -(D-4) \frac{B}{r^2}
 + \c \phi'\frac{3B}{r}\Big], \\
\label{tilh-def}
\tilde h &=& 1+2(D-3) e^{-\c\phi} \Big[-(D-4)\frac{B}{r^2}
+\c\phi'\frac{2B}{r} \Big],
\ena
\vs{-5}
\bea
\label{equ}
U(r) &=& (2 \tilde h)^{-1} \Bigg[ - \frac{(D-3)_4}{r^2}
-2\frac{D-3}{r}\Big(\frac{B'}{B}-\d'\Big) -\frac12 \phi'^2
+ (D-3)e^{-\c\phi} \Bigg\{ (D-4)_6 \frac{ B}{r^4} \nn
&& \hs{10} + 4 (D-4)_5 \frac{B}{r^3}\Big(\frac{B'}{B}-\d'-\c\phi'\Big)
+4(D-4)\c \frac{B}{r^2}\Big( \c \phi'^2 +\frac{D-2}{r}\phi'-\Phi \Big) \nn
&& \hs{10} +8 \frac{\c\phi'}{r} \biggl[\Big(\frac{B'}{2}-\d' B\Big)\Big(\c\phi'-\d'
+\frac{2}{r} \Big) -\frac{D-4}{2r}B' \biggr]
+ 4(D-4)\Big(\frac{B'}{2B}-\d' \Big) \frac{B'}{r^2}
-\frac{4\c}{r}\Phi (B'-2\d'B)\Bigg\} \nn
&& \hs{10} +\frac{1}{2B}e^{2\d-\c\phi} f'^2
+\frac{3}{16} B \phi'^4 e^{-\c\phi}
-\frac{1}{B} {\Lambda}e^{\lambda\phi}\Biggr],
\\
\Phi &=& \phi'' +\Big(\frac{B'}{B}-\d' +\frac{D-2}{r}\Big) \phi'.
\label{dil}
\ena
The original field equations involve the function $U(r)$ defined by
\bea
U(r) \equiv \Big(\frac{B'}{2B} -\d' \Big)' +\Big(\frac{B'}{2B} -\d' \Big)^2
+ \frac{B'}{2B}\Big(\frac{B'}{2B} -\d' \Big).
\label{originalu}
\ena
However in the field equations we have two equations involving $\d''$,
one of which can be used to eliminate $\d''$ from $U(r)$ and obtain the expression~\p{equ}.
The resulting equations are the above set of equations with only the first derivative of $\d$.

The field equation for the Maxwell field~\p{fe4} is easily integrated to give
\bea
f'=\frac{q}{r^{D-2}} e^{\c\phi-\d},
\label{charge}
\ena
where $q$ is a constant corresponding to the charge.
So we can simply substitute this into our basic equations, and do not have to
integrate the field equation for the Maxwell field simultaneously.

Our task is now reduced to setting boundary conditions for the metric functions $B,\: \d$ and
the dilaton field $\phi$, and integrate the above set of equations~\cite{GOT1,GOT2,OT3,OT4,OT5}.

\subsection{Symmetry and scaling}

Before proceeding to the numerical analysis,
it is useful to consider several symmetries of our field equations
(or our model).
First, our field equations~\p{fe1}-\p{fe3} are
invariant under the scaling transformation
\bea
B \to a^2 B,~~
r \to a r,~~
q \to a^{D-2} q,
\label{sym1}
\ena
with an arbitrary nonzero constant $a$.
If a black hole solution with the horizon radius $r_H$ is obtained,
we can generate solutions with different horizon radii and charge but the same $\Lambda$
by this scaling transformation.

Second, there is the shift symmetry under
\bea
\phi \to \phi-\phi_{\ast}, ~~
\Lambda \to e^{(\la-\c)\phi_{\ast}} \Lambda,~~
B \to e^{-\gamma\phi_{\ast}}B,
\label{sym2}
\ena
where $\phi_{\ast}$ is an arbitrary constant.
This changes the magnitude of the cosmological constant. Hence, given a solution
with some value of the cosmological constant, this symmetry may be used to generate
solutions for different values of the cosmological constant.

The final one is another shift symmetry under
\bea
\delta \to \delta - \delta_{\ast}, ~~
t \to  e^{-\delta_{\ast}}t,
\label{sym3}
\ena
with an arbitrary constant $\delta_{\ast}$, which may be used to shift
the asymptotic value of $\delta$ to zero.

We will use these symmetries to obtain general black hole solutions after constructing
solutions for given charge and cosmological constant.

\subsection{Boundary conditions and asymptotic behaviors}

In this paper we consider nonextreme solutions which have a nondegenerate black hole horizon.
At the horizon $r_H$, we impose the conditions
\bea
B_H=0,~~
B_H' \neq 0,
\ena
where $B_H=B(r_H)$.
Here and in what follows, quantities evaluated at the horizon are represented by a subscript $H$.
At the horizon, Eqs.~\p{fe1}--\p{fe4} together with \p{h-def}--\p{charge} give
\bea
B_H' &=& -\frac{1}{D-2} \Big[ r_H \Lambda e^{\la\phi_H}
+\frac{q^2}{2 r_H^{2D-5}}e^{\c\phi_H}\Big],
\label{boundary_B}
\\
\phi_H' &=& 2\c\frac{(D-2)_3}{r_H^2}e^{-\c\phi_H} B_H'
 +\frac{1}{B_H'} \Big[ \la\Lambda e^{\la\phi_H}+\frac{\c q^2}{2r_H^{2D-4}}e^{\c\phi_H}\Big],
\label{boundary_phi}
 \\
\d_H' &=& -\frac{1}{2(D-2)}r_H(\phi_H')^2.
\label{boundary_delta}
\ena

In the asymptotic region of $r\to \infty$, we assume
\bea
\label{bc1}
B &\sim& b_2 r^2 - \frac{2M}{r^\mu}+\cdots, \nn
\delta &\sim& \d_0 +\frac{\d_1}{r^\s} +\cdots, \\
\phi &\sim& \phi_0 +\frac{\phi_1}{r^\nu} + \cdots, \nonumber
\ena
with finite constants $b_2>0$, $M$, $\d_0$, $\d_1$, $\phi_0$, $\phi_1$ and positive constants
$\mu$, $\s$, $\nu$; $b_2$ is related to the AdS radius.

In order to understand the asymptotic behaviors of our dilatonic system, it is convenient
to study the ``effective potential'' for the dilaton. Its field equation can be written as
\bea
\squaret \phi=\frac{dV_{\rm eff}}{d\phi},
\ena
where the effective potential is defined by
\bea
V_{\rm eff} = -e^{-\c\phi} \Big[ R_{\rm GB}^2+\frac{3}{16}(\pa\phi)^4-\frac{1}{4}F^2 \Big]
+\Lambda e^{\la\phi}.
\label{dpot}
\ena
A similar effective potential was analyzed in our earlier paper~\cite{GOT2} for
neutral AdS black holes in the absence of the Maxwell field and higher derivative term
in the dilaton. Actually these additional contributions do not change the asymptotic
behavior and we can take the results there. According to that analysis,
we can have asymptotically AdS solutions only for $\la>0$.
(The basic reason is that the potential~\p{dpot} must have a maximum
to which the dilaton $\phi$ tends at $r\to\infty$, and this is possible only for $\la>0$.)
So hereafter we restrict our analysis to this case.

\subsection{Asymptotic expansion}

We now analyze the asymptotic behaviors of the solutions (see Ref.~\cite{GOT2}).
Substituting the asymptotic behaviors~\p{bc1} into the field equations~\p{fe1} and \p{fe3}
and using the original definition \p{originalu} of the function $U$,
we require that the leading terms (proportional to $r^2$) should vanish. This gives
\bea
\label{ase1}
&& (D-1)_4 e^{-\c\phi_0} b_2^2-(D-1)_2 b_2 -\Lambda e^{\la\phi_0} =0, \\
&& (D)_3 \c e^{-\gamma \phi_0}b_2^2 +\la\Lambda e^{\la\phi_0}=0,
\label{ase2}
\ena
which determine $b_2$ and $\phi_0$, while $\d_0$ can be arbitrary due to the shift
symmetry~\p{sym3}. Since $\c$ is positive and $\Lambda$ is negative, it follows from \p{ase2}
that $\la$ should be positive in accordance to what we stated above.
These are the same set of equations as in the neutral case~\cite{GOT2},
and we can take the results from there. There we find that there are sensible solutions
for $\la\neq\c$, to which we restrict our following discussions.
Then Eqs.~\p{ase1} and \p{ase2} give
\bea
b_2^2 &=& -\frac{\la\Lambda}{(D)_3 \c} \Big[\frac{D(D-3)}{(D-1)_2} (-\Lambda)\frac{\c}{\la}
\Big(1+\frac{(D-4)\la}{D\c} \Big)^2 \Big]^{(\c+\la)/(\c-\la)}, \\
e^{\phi_0} &=& \Big[ \frac{D(D-3)}{(D-1)_2}(-\Lambda)\frac{\c}{\la}
\Big(1+\frac{(D-4)\la}{D\c} \Big)^2 \Big]^{1/(\c-\la)}.
\label{phi_0}
\ena

We find that the contributions from the field strength of the Maxwell field and
the higher order dilaton terms are also subleading compared to other terms present
in the neutral case, so the analysis of the next leading terms in Eqs.~\p{fe1}--\p{fe3}
is again completely the same as in Ref.~\cite{GOT2}. Now let us summarize the results obtained there.
First we have
\bea
(D-4)\la-D\c \neq 0,~~~~
\mu=\nu-2=\s-2.
\ena
The asymptotic forms of the field functions are given by
\bea
\label{ase3}
B &\sim& b_2 r^2 -\frac{2M_+}{r^{\nu_+-2}} -\frac{2M_0}{r^{D-3}} + \cdots, \nn
\d &\sim& \frac{\d_+}{r^{\nu_+}} + \cdots, \\
\phi &\sim& \phi_0 + \frac{\phi_+}{r^{\nu_+}} + \cdots, \nonumber
\ena
where we have set $\d_0=0$ using the shift symmetry~\p{sym3} and have defined
\bea
\nu_\pm = \frac{D-1}{2} \left( 1 \pm \sqrt{1-\frac{m^2}{m_{BF}^2}}\; \right),
\ena
and
\bea
m_{BF}^2 \equiv -\frac{(D-1)^2}{4} b_2,~~~
m^2 \equiv -\frac{(D)_2\la\c(\la-\c)[(D-4)\la+D\c]}{(D-4)^2\la^2-D^2\c^2-8(D-1)_2\la^2 \c^2}
b_2.
\ena
The mass square $m_{BF}^2$ is that of Breitenlohner and Freedman (BF) bound~\cite{BF}.
In the asymptotic expansion of $\phi$ in \p{ase3}, there would be a term $1/r^{\nu_-}$
in general, but we choose the boundary value $\phi_H$ such that this term is absent.

The function $B$ has the term $1/r^{\nu_+ -2}$, but it can be shown that this term
cancels out from the $g_{tt}$ component of the metric due to the condition that
the next leading terms in the field equations cancel out (we have $M_+ + b_2 \d_+=0$ from
that condition, see Eq.~(3.23) in Ref.~\cite{GOT2}).
This leads to
\bea
g_{tt} = B e^{-2\d} \sim b_2 r^2 -\frac{2M_0}{r^{D-3}} + \cdots.
\ena
This behavior is also confirmed by our numerical analysis.
The value $M_0$ thus gives the gravitational mass of the black hole.
Consequently it is convenient to define the mass function $m_g(r)$ by
\bea
-g_{tt} = b_2 r^2 -\frac{2m_g(r)}{r^{D-3}}.
\ena
We also have to impose the conditions
\bea
m^2<0,
~~~\mbox{and}~~~
m^2 \geq m_{BF}^2,
\label{cond}
\ena
in order for our black hole solutions to exist and be stable.
The first condition gives the allowed parameter regions
\bea
\la<\c,
\ena
or
\bea
\la> \frac{D\c}{\sqrt{(D-4)^2 - 8(D-1)_2 \c^2}}, ~~~
\mbox{and}~~~
0<\c< \frac{D-4}{\sqrt{8(D-1)_2}}.
\ena
In constructing our black hole solutions, we choose the parameters in these regions.

Here we briefly comment on the extreme black hole solutions.
At the degenerate horizon, where $B'_H=0$, we have
\bea
r_H \Lambda e^{\la\phi_H}
+\frac{q^2}{2 r_H^{2D-5}}e^{\c\phi_H}=0,
\label{ext1}
\ena
by Eq.~(\ref{boundary_B}), and
\bea
\la\Lambda e^{\la\phi_H}+\frac{\c q^2}{2r_H^{2D-4}}e^{\c\phi_H}=0,
\label{ext2}
\ena
by the regularity of $\phi'_H$ in Eq.~(\ref{boundary_phi}).
These equations give the conditions
\bea
\gamma=\lambda,
~~~~~
\Lambda+\frac{q^2}{2 r_H^{2D-4}}=0.
\label{ext3}
\ena
When $\gamma=\lambda$, Eqs.~(\ref{ase1}) and (\ref{ase2}) give
\bea
\Lambda=-\frac{(D)_1}{4(D-2)_3}.
\ena
These are the necessary conditions on the existence of the extreme solution.
We thus see that the existence of the extreme black hole solutions depends on
the very specific choice of parameters in our system and is not generic.
Unless we make such a choice, we cannot have extreme solutions.

Let us also comment on the charge of the Maxwell field. By the existence of the cosmological
constant $\Lambda$, the asymptotic value of the dilaton field is not zero but takes nonzero
finite value $\phi_0$ determined by Eq.~(\ref{phi_0}). This changes the coefficient
of the field strength $f'$ at infinity as $q\to qe^{-\gamma \phi_0}$. This means
that the effective charge of the solution is $q_{\rm eff}=qe^{-\gamma \phi_0}$.
We treat, however, the $q$ as the parameter of the solution in the following analysis.

\subsection{Thermodynamical variables}

The Hawking temperature is given by the periodicity of the Euclidean time
on the horizon as
\bea 
\label{temp}
T_H \!\!\!\!\! && =\frac{e^{-\d_H}}{4\pi}B_H'
\nonumber \\
&& =-\frac{e^{-\d_H}}{4\pi(D-2)}
\biggl[
r_H \Lambda e^{\la\phi_H}+\frac{q^2e^{\c\phi_H}}{2r_H^{2D-5}}
\biggr].
\label{temperature}
\ena 
Note that the cosmological constant is negative $\Lambda<0$, and the temperature becomes positive.
Also this vanishes if extreme condition is satisfied.

Using the definition of entropy in the diffeomorphism invariant system in Ref.~\cite{Wald},
we obtain
\bea 
S=-2\pi \int_\Sigma \frac{\pa {\cal L}}{\pa R_{\mu\nu\rho\sigma}}
\e_{\mu\nu} \e_{\rho\sigma},
\ena 
where $\Sigma$ is the event horizon $(D-2)$-surface, ${\cal L}$ is the Lagrangian
density, $\e_{\mu\nu}$ denotes the volume element binormal to $\Sigma$.
This entropy has desirable properties such that it obeys the first law of black
hole thermodynamics and that it is expected to obey even the second law~\cite{Jacobson}.
For our present model, this gives
\bea 
S = \frac{r_H^{D-2}\Sigma_{0}}{4}.
\label{entropy}
\ena 
$\Sigma_0$ is the volume of the unit constant curvature space.
This is the same form as the Bekenstein-Hawking entropy, i.e., a quarter of the horizon area.

%

\section{Numerical solutions of the dilatonic black holes}
\label{sec3}

\subsection{Parameters and scaling symmetry}

We construct the black hole solutions numerically.
For this purpose, we have to choose the parameters for our black hole solutions
so as to satisfy the condition~\p{cond}.
We choose the following values as a typical example in various dimensions:
\bea
\c=\frac12,~~~
\la=\frac13,~~~
\d_0=0,~~~
\Lambda=-1,~~~
q=1,~~~
\phi_- =0.
\label{para}
\ena
The last condition $\phi_-=0$ means that the asymptotic behavior of $\phi$ in \p{ase3}
is chosen such that the term like $1/r^{\nu_-}$ is absent.
This is achieved by choosing suitable $\phi_H$.
It should not be difficult to get solutions for other values of parameters
in the allowed region.

We next fix the radius of the event horizon $r_H$,
and choose the value of the dilaton field $\phi_H$ at the horizon,
and determine the other fields by \p{boundary_B}-\p{boundary_delta}.
Once a solution for one $r_H$, mass and charge is found,
we can use the symmetry~\p{sym1} to obtain other solutions by changing
\bea
r_H \to ar_H,~~
M_0 \to a^{D-1} M_0,~~
q\to a^{D-2} q,
\label{scaling_1}
\ena
where $a$ is a constant parameter.
Similarly, given a solution for a cosmological constant, we can generate solutions for different
cosmological constants but the same $r_H$ and $q$ using transformation~\p{sym2}:
\bea
M_0 \to b^{2\c/(\c-\la)}M_0,~~
\Lambda \to b^2 \Lambda,~~
\phi \to \phi+\frac{2}{\c-\la} \ln b,
\label{scaling_2}
\ena
with a constant parameter $b$.

\subsection{Neutral black holes}
\label{secd=neutral}

Before proceeding to the charged solution which is the main issue of the paper,
let us comment on the neutral solutions. Because the present system (\ref{action1})
includes the higher order term of the dilaton field $(\partial \phi)^4$,
which is neglected in the previous analysis in Refs.~\cite{GOT1,GOT2,OT3,OT4,OT5},
we should check how different our solutions may be from those without the higher order term.
In the neutral case the following relation was derived by combining the symmetries
(\ref{scaling_1}) and (\ref{scaling_2}):
\begin{equation}
M_0 = m_0 |\Lambda|^{\c/(\c-\la)}\:r_H^{\:D-1},
\label{neutm}
\end{equation}
where $m_0$ is a coefficient.

We have constructed the neutral solutions numerically with the higher derivative term
$(\pa\phi)^4$ included.
Here we summarize the obtained values of the physical parameters in Table~\ref{table_neutral}
for $D=4,\:5,\:6$, and 10. Since the solutions are qualitatively similar for dimensions
beyond six, we do not give them explicitly, but
$D=10$ is the critical dimension in superstring theories and is important for applications.
The word ``on" (``off") in the column of $(\partial \phi)^4$ means that the higher order
term of the dilaton field is (not) included.
The values for ``off'' are those obtained in our previous paper~\cite{GOT2}.
It can be seen that the effect of the higher
order term is tiny. The difference in the values of physical quantities becomes smaller
as the dimension of the spacetime $D$ gets higher, and are less than 1\% for all parameters
in each dimensions.
Hence we can conclude that the analysis in Ref.~\cite{GOT2} is valid with enough accuracy.
\begin{table}[tbh]
\caption{
Values of the physical parameters of the neutral black hole solutions obtained by numerical
calculations.
The rows with $(\partial \phi)^4$ ``on" (``off") means that the higher order term of
the dilaton field is (not) included. ``off'' case is obtained in \cite{GOT2}.
}
\begin{tabular}{c|c|c|c|c|c}
\hline
~~~$D$~~~ &
~~~$(\partial \phi)^4$~~~ &
\hspace{8mm} $\phi_H$ \hspace{8mm} &
\hspace{8mm} $\delta_H$ \hspace{8mm} &
\hspace{8mm} $m_0$ \hspace{8mm} &
\hspace{8mm} $T$ \hspace{8mm} \\
\hline\hline
4 & off & $-$0.098566425~ & $-$0.028926580 & 0.08300 & 0.039632717   \\
\cline{2-6}
   &on  & $-$0.098650689~ & $-$0.028953439 & 0.08300 & 0.039632668   \\
\hline
5 & off & \:\:2.767015400 & $-$0.021880831 & 0.14006 & 0.068192735  \\
\cline{2-6}
   &on  & \:\:2.766998583 & $-$0.021886221 & 0.14006 & 0.068192721   \\
\hline
6 & off & \:\:4.154180204 & $-$0.016209541 & 0.15948 & 0.080751210   \\
\cline{2-6}
   &on  & \:\:4.154175817 & $-$0.016210949 & 0.15948 & 0.080751206   \\
\hline
10 & off & \:\:6.291642457 & $-$0.002453368 & 0.13216 & 0.081203391  \\
\cline{2-6}
     & on & \:\:6.291639521 & $-$0.002453809 & 0.13216 & 0.081203348    \\
\hline
\end{tabular}
\label{table_neutral}
\end{table}

\subsection{Charged black holes}
\label{sec_charged}

We now turn to the charged solutions.
We fix the parameters as $r_H=1$, $q=1$ and $\Lambda=-1$.
We then integrate the field equations~\p{fe1}-\p{fe3}
outward from the horizon numerically with the additional
conditions~ \p{boundary_B}-\p{boundary_delta}.
By integrating the field equations with the value $\phi_H$, we have found that the dilaton
field increases monotonically to $\phi_0$ at infinity.

With the parameters~\p{para}, we have constructed the black hole solutions and found  $b_2$
and the asymptotic value of the dilaton field $\phi_0$.
They are summarized in Table~\ref{table1}.
These values do not depend on the charge $q$.
For $D=4$, the asymptotic value of the dilaton field vanishes.

\begin{table}[tbh]
\caption{The AdS curvature radius $b_2$ and the asymptotic value of the dilaton field
$\phi_0$ for the parameter $\c=\frac12$, $\la=\frac13$, $\Lambda=-1$ in each dimension.}
\begin{tabular}{c|c|c}
\hline
~~~$D$~~~ & ~~~~~~$b_2$~~~~~~ & ~~~~~~$\phi_0$~~~~~~  \\
\hline\hline
4 & 0.16667  & 0~~~~~~~\:\:   \\
\hline
5 & 1.50205  & 0.29147   \\
\hline
6 & 0.24346  & 2.84082   \\
\hline
10 & 0.15886  & 6.30143   \\
\hline
\end{tabular}
\label{table1}
\end{table}

\begin{figure}[htb]
\begin{center}
\includegraphics[width=12cm]{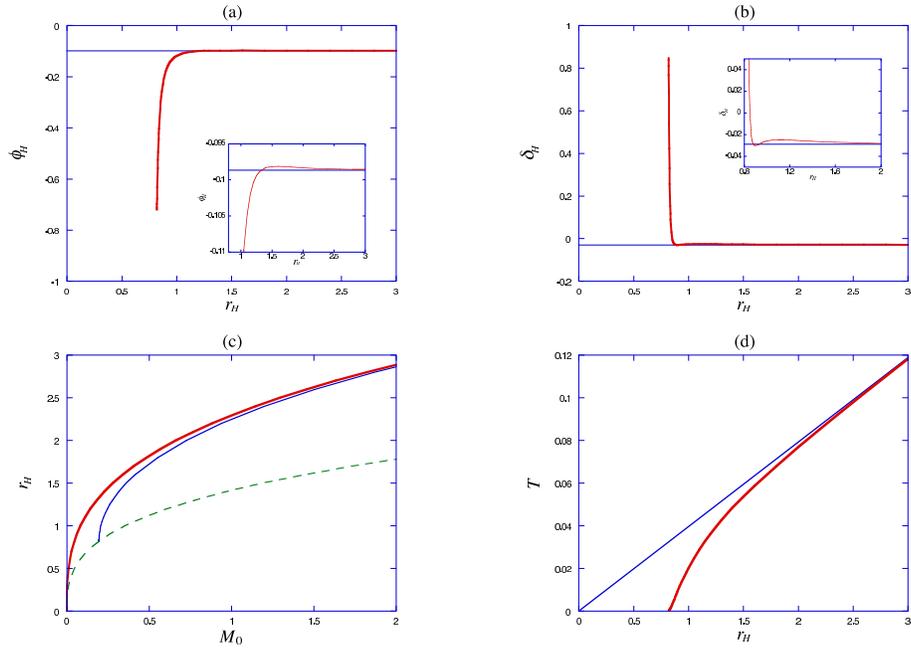}
\end{center}
\vspace*{-5mm}
\caption{
Various relations between the parameters for $D=4$ black hole solutions in (dilatonic)
EMGB systems.
(a) $r_H$-$\phi_H$ diagram (b) $r_H$-$\delta_H$ diagram (c) $M_0$-$r_H$ diagram (d) $r_H$-$T$
diagram. Thick (red) lines for the charged solutions with $q=1$ and thin (blue)
lines for the neutral solutions.
Below the dashed (green) curve in (c), there is no solution for any $q$.
}
\label{d4-1}
\end{figure}

\begin{figure}[hbt]
\begin{center}
\includegraphics[width=12cm]{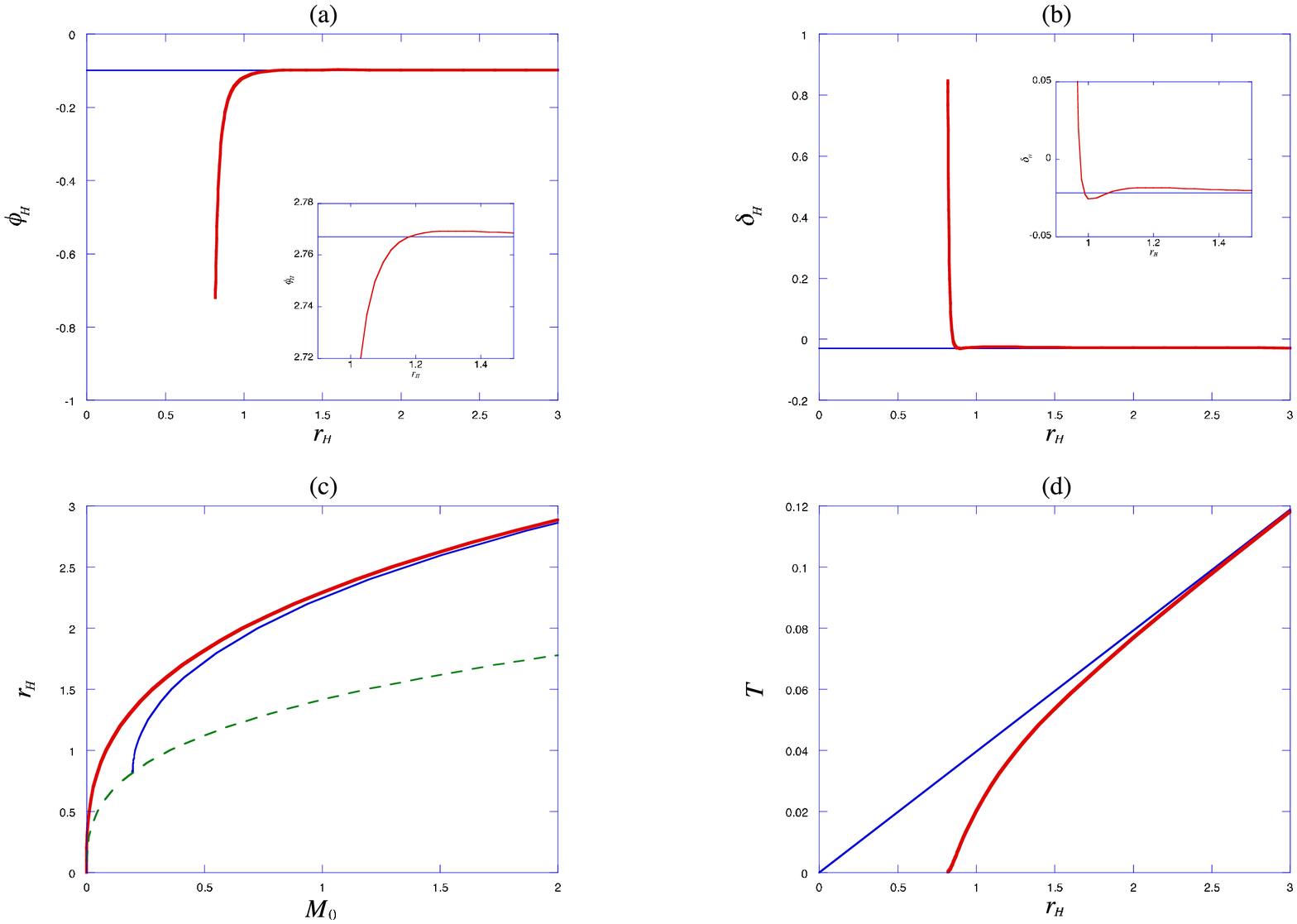}
\end{center}
\vspace*{-5mm}
\caption{
Various relations between the parameters for $D=5$ black hole solutions in (dilatonic)
EMGB systems.
(a) $r_H$-$\phi_H$ diagram (b) $r_H$-$\delta_H$ diagram (c) $M_0$-$r_H$ diagram (d) $r_H$-$T$
diagram. Thick (red) lines for the charged solutions with $q=1$ and thin (blue) lines
for the neutral solutions.
Below the dashed (green) curve in (c), there is no solution for any $q$.
}
\label{d5-1}
\end{figure}

\begin{figure}[hbt]
\begin{center}
\includegraphics[width=12cm]{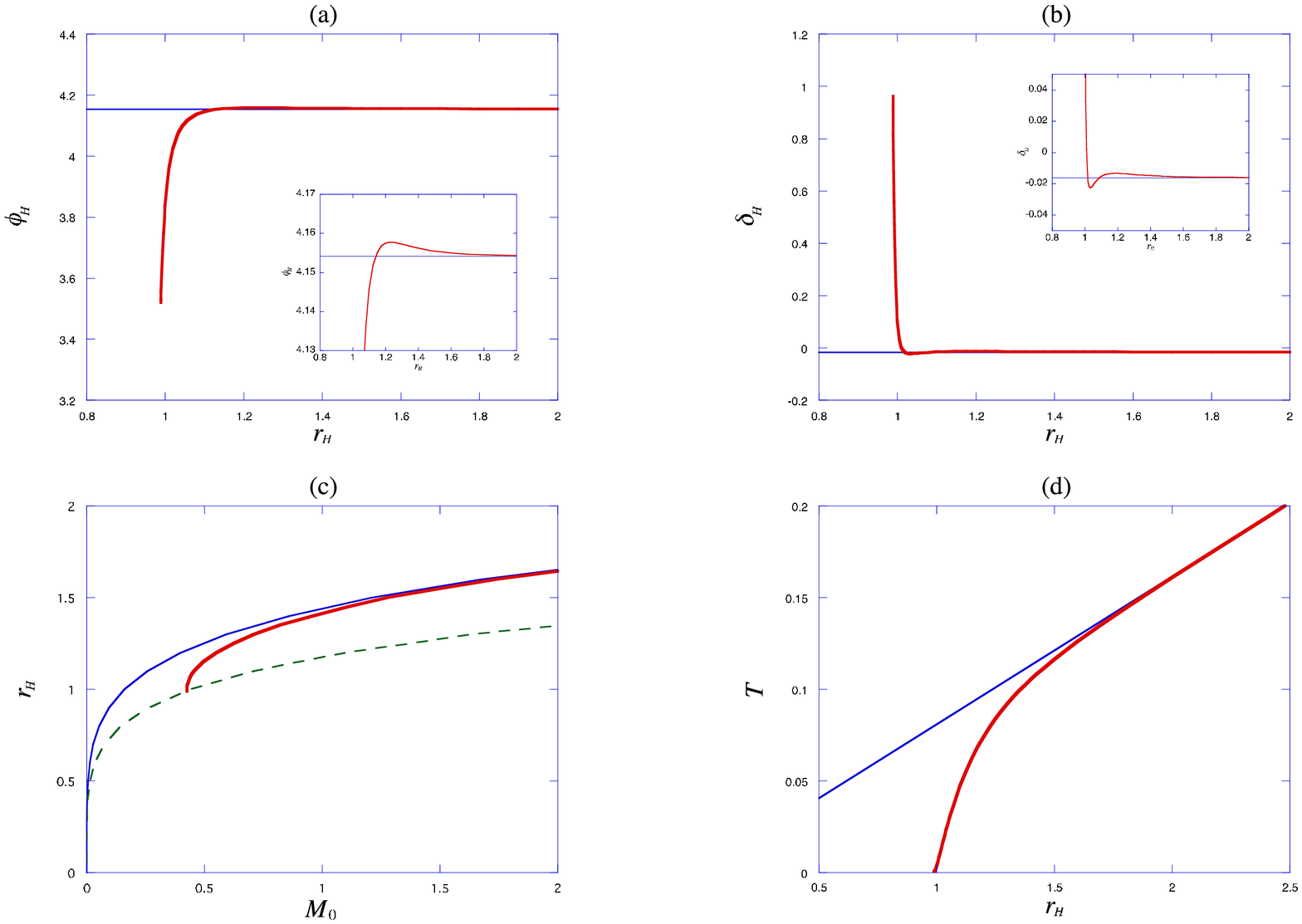}
\end{center}
\vspace*{-5mm}
\caption{
Various relations between the parameters for $D=6$ black hole solutions in (dilatonic)
EMGB systems.
(a) $r_H$-$\phi_H$ diagram (b) $r_H$-$\delta_H$ diagram (c) $M_0$-$r_H$ diagram
(d) $r_H$-$T$ diagram.
Thick (red) lines for the charged solutions with $q=1$ and thin (blue) lines for the neutral
solutions.
Below the dashed (green) curve in (c), there is no solution for any $q$.
}
\label{d6-1}
\end{figure}

\begin{figure}[htb]
\begin{center}
\includegraphics[width=12cm]{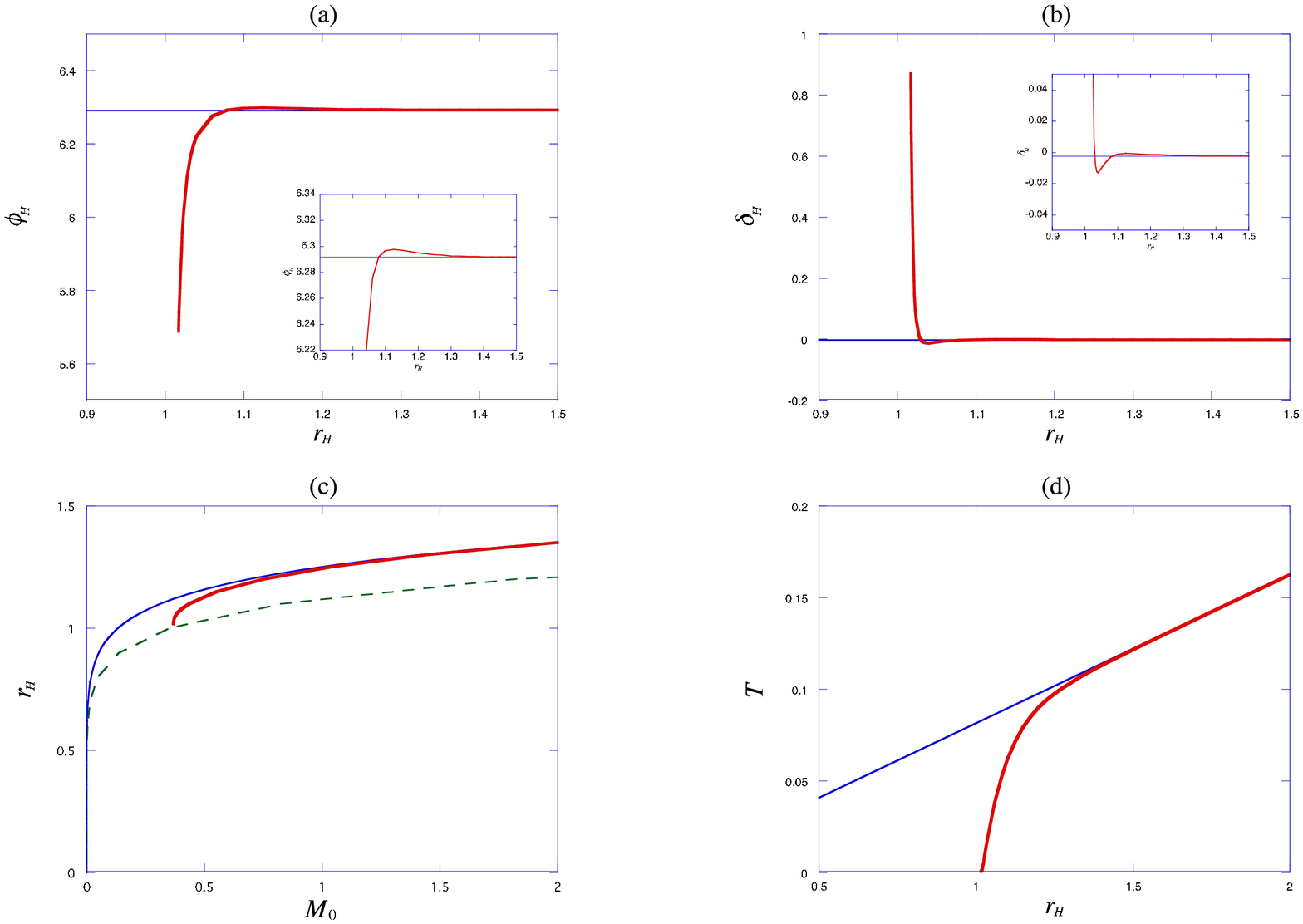}
\end{center}
\vspace*{-5mm}
\caption{
Various relations between the parameters for $D=10$ black hole solutions in (dilatonic)
EMGB systems.
(a) $r_H$-$\phi_H$ diagram (b) $r_H$-$\delta_H$ diagram (c) $M_0$-$r_H$ diagram (d) $r_H$-$T$
diagram. Thick (red) lines for the charged solutions with $q=1$ and thin (blue) lines
for the neutral solutions.
Below the dashed (green) curve in (c), there is no solution for any $q$.
}
\label{d10-1}
\end{figure}

Figures~\ref{d4-1}--\ref{d10-1} show the relations $r_H$-$\phi_H$, $r_H$-$\delta_H$, $M_0$-$r_H$,
and $r_H$-$T$ of the obtained black hole solutions in $D=4$, 5, 6 and 10, respectively.
The solutions are shown for the charge $q=1$.
By the scaling symmetries, these relations do not change if we make replacement
$r_H \to r_H/q^{\frac{1}{D-2}}$, $M_0 \to M_0/(|\Lambda|^3 q^{\frac{D-1}{D-2}})$,
$T \to T/(|\Lambda|^3 q^{\frac{1}{D-2}})$, $\phi_H \to\phi_H-6\ln |\Lambda|$.
$\delta_H$ is kept invariant under the symmetries.
What this means is that if we replace the variables by this rule, these figures remain
the same and we can get the relations among physical quantities including mass,
horizon radius and temperature for the black hole solutions with other
values of charges and cosmological constants.
Those of the neutral solutions are also plotted for comparison.
In particular, in diagrams (a) and (b) in Figs.~\ref{d4-1}--\ref{d10-1},
the blue horizontal lines show the values of $\phi_H$ and $\delta_H$ which are constant
for neutral solutions and are given in Table~\ref{table_neutral}.
Their values for charged solutions asymptotically approach these values for neutral solutions
because the effects of the charges become smaller for larger horizon radius $r_H$.
To show precisely how $\phi_H$ and $\delta_H$ behaves near the horizontal line,
we insert enlarged figures around there as an inset.

Several features can be read off from the diagrams.
Firstly we see that the qualitative behaviors of the diagrams do not depend on
the dimensions $D$ much.
For the neutral solutions, there are infinitesimally small black hole solutions in
all dimensions as the horizon radius becomes small. This can be seen from
the relation~\p{neutm} in this case.
The horizon becomes degenerate ($B_H' \to 0$) and the temperature approaches zero
in the limit of zero horizon radius.
For the charged black holes away from the small horizon radius, mass and temperature
similarly become smaller for smaller horizon.
However, the situation drastically changes near small horizon radius
and there is no such limit of zero horizon radius.
There is nonzero lower limit for the radius of the event horizon below
which no solution exists like Reissner-Nordstr\"om solution.
However, the solution in the lower limit is not the extreme solution
because it can exist only for $\lambda=\c$ as shown in
\p{ext3}, which is not satisfied here.
The lower limits are found to be $r_H^{\ast}=0.8201,\; 0.9491,\; 0.9891,\; 1.0173$
for $D=4,\;5,\;6,\;10$, respectively.
Around these limits, the numerical calculation becomes unstable and it is difficult
to say anything definite about the nature of this lower limit.

For $D=4$, $\phi_H$ is negative for all $r_H$, while it is positive for $D\geq5$.
$\phi_H$ becomes small rapidly as the horizon radius of the solution approaches the lower limit.
On the other hand, $\delta_H$ increases sharply towards this limit.
For the mass of the black hole, there is the lower limit
$M_0^{\ast}=0.1940,\; 0.360,\; 0.426,\; 0.368$ for $D=4,\;5,\;6,\;10$, respectively.
{}From the data of the lower limit and the scaling symmetries, we find that the curve
connecting the lower limits for each $q$ is
\begin{equation}
M_0^{\ast} = m^{\ast} |\Lambda|^{\gamma/(\gamma-\lambda)} \:{r_H^{\ast}}^{D-1}
~~~
(\;=m^{\ast} |\Lambda|^3 \:{r_H^{\ast}}^{D-1}~ \mbox{ for the choice~\p{para}}).
\end{equation}
where $m^{\ast} = 0.376,\; 0.421,\; 0.440,\; 0.350$ for $D=4,\;5,\;6,\;10$, respectively.
This is plotted by the dashed curves in the $M_0$-$r_H$ diagrams for any $q$.
Below these curves there is no black hole solution.
While the temperature seems to be zero in the lower limit in the $r_H$-$T$ diagrams,
it is not the case.
In fact the solutions at the lower limit have tiny but nonzero temperature
$T^{\ast}=2.157\times 10^{-4},\; 4.413\times 10^{-4},\; 4.520\times 10^{-4},\;
 5.480\times 10^{-4}$ for $D=4,\;5,\;6,\;10$, respectively.

It is expected that the difference between charged and neutral solutions becomes
small for large black holes, and this is confirmed in our solutions.

We have discussed qualitative properties of the black hole solutions.
To evaluate actually physical quantities, it is necessary to have quantitative results.
In order to get idea on what are the typical physical quantities,
here we tabulate them  for the black hole solutions in Table \ref{table_parameter}.
The parameters change rapidly around the lower limit.
\begin{table}[h]
\caption{Typical values of the physical quantities of the charged dilatonic black hole
solutions for $\Lambda=-1$, $q=1$.
}
\begin{tabular}{@{~~~}c@{~~~}|l@{~~}|@{~~~}l@{~}|@{~~~}l@{~~}|@{~~}l@{~~}|@{~~}l@{~~}|@{~~~}l@{~~~}}
\hline
$D$ & ~~~~$r_H$ & ~~$M_0$ & ~~~~~$\delta_H$ & ~~~~\:$\phi_H$ & ~~~~$\phi_H'$ & ~~~~~~$T$
  \\
\hline\hline
4 & ~~0.8201 & 0.194~~ & ~\:\:0.8469  &  $-$0.7198 & 19.54  &$2.157\times 10^{-4}$     \\
\cline{2-7}
 & ~~0.9 &  0.197 &  $-$0.03054 & $-$0.1815  & ~\:0.6084 &  0.009058  \\
\cline{2-7}
 & ~~1.0 & 0.205  & $-$0.02611  & $-$0.1097  & ~\:0.1584  &  0.02001  \\
\cline{2-7}
 & ~~1.5 & 0.359  & $-$0.02697  & $-$0.09817  & ~\:0.1913  & 0.05355   \\
\cline{2-7}
 & ~~2.0 &  0.725 & $-$0.02826  & $-$0.09835  & ~\:0.1646  & 0.07678    \\
\hline
5 & ~~0.9491 & 0.360  & ~\:\:0.8358 &  ~\:\:2.159 & 25.21  &  $4.413\times 10^{-4}$   \\
\cline{2-7}
 & ~~1.0 & 0.364  & $-$0.02594  &  ~\:\:2.671 & ~\:1.058  & 0.01456    \\
\cline{2-7}
 & ~~1.2 & 0.446  & $-$0.01857  & ~\:\:2.768  &  ~\:0.08905 &   0.05991  \\
\cline{2-7}
 & ~~1.5 & 0.810  &  $-$0.02049 &  ~\:\:2.768 &  ~\:0.1403  & 0.09508     \\
\cline{2-7}
 & ~~2.0 & 2.298  & $-$0.02161  & ~\:\:2.767  & ~\:0.1258  & 0.13467    \\
\hline
6 & ~~0.9891 & 0.426  & ~\:\:0.9610  & ~\:\:3.520  & 34.30  & $4.520\times 10^{-4}$   \\
\cline{2-7}
 & ~~1.0 &  0.426 & ~\:\:0.1062  &  ~\:\:3.834  & 10.73  & 0.003386    \\
\cline{2-7}
 & ~~1.2 &  0.55 & $-$0.01345  & ~\:\:4.157  &  ~\:0.01678 & 0.07425    \\
\cline{2-7}
 & ~~1.4 &  0.96 & $-$0.01495  & ~\:\:4.156  &  ~\:0.08153 & 0.1053    \\
\cline{2-7}
 & ~~1.6 &  1.74 & $-$0.01574  & ~\:\:4.155  & ~\:0.09288  &  0.1262   \\
\cline{2-7}
 & ~~1.8 &  3.06 & $-$0.01602  & ~\:\:4.154  &  ~\:0.08903 & 0.1440    \\
\hline
10 & ~~1.0173 & 0.368  & ~\:\:0.8703  & ~\:\:5.688  & 63.624  & $5.480\times 10^{-4}$   \\
\cline{2-7}
& ~~1.1 &  0.438 & $-$0.001355  &  ~\:\:6.297  & \!\!$-$0.3420  & 0.06160    \\
\cline{2-7}
 & ~~1.15 &  0.552 & $-$0.0008287  & ~\:\:6.297  &  \!\!$-$0.3041 & 0.07914    \\
\cline{2-7}
 & ~~1.2 &  0.75 & $-$0.001380  & ~\:\:6.295  & \!\!$-$0.2440  &  0.08992   \\
\cline{2-7}
 & ~~1.3 &  1.44 & $-$0.002105  & ~\:\:6.293  & \!\!$-$0.1827  &  0.10330   \\
\cline{2-7}
 & ~~1.4 &  2.76 & $-$0.002343  & ~\:\:6.292  &  \!\!$-$0.1579 & 0.1129    \\
\hline
\end{tabular}
\label{table_parameter}
\end{table}

\section{Conclusions 
}
\label{CD}

In this paper we have studied asymptotically AdS charged black hole solutions
in the dilatonic EMGB theory including the cosmological constant in various dimensions.
The theory originates from the low-energy effective theory of the heterotic string.
The spacetime is assumed to be static and plane symmetric, and should be regular
outside of the black hole horizon.
The system of the field equations has some symmetries which are used to obtain
solutions with different masses and charges.

As the inner boundary conditions, we have assumed the existence of the regular black hole
horizon. At infinity, the dilaton field decays fast enough for satisfying the BF bound for
the stability. By these conditions, it is shown that the extreme black hole solution with
degenerate horizon can exist only when the theoretical parameters satisfy very stringent
conditions, and the existence of these is non-generic in the parameter space.
Hence we only consider the solutions with non-degenerate horizon.
The system of the field equations is complex so it is difficult to
obtain an analytical solution. Hence we have investigated the solutions numerically.

Since the system in this paper includes the higher order term of the dilaton field,
we have also analyzed the neutral solutions to see the difference from the solutions
we obtained in Ref.~\cite{GOT2}, which do not include such term. The numerical analysis
shows that there is only difference less than 1\% in the physical parameters between
these solutions. The higher order term of the dilaton field does not much affect the solutions.

For the charged solutions, we have obtained the relations between the physical quantities
(mass, temperature and horizon radius) for $D=4, 5, 6$ and 10.
It is found that the properties of the solutions are qualitatively the same independently
of the dimensions.
We find that the mass and temperature become smaller for smaller horizon.
However, there is a nonzero lower limit for the radius of the event horizon below which
no solution exists. This is in sharp contrast to the neutral black holes where the horizon
radius can go to zero together with the mass and temperature.
We have shown the regions where no solution exists in the mass-horizon radius diagram.
We have also confirmed that the difference between charged and neutral solutions becomes
smaller for larger black holes.

With these solutions, we should be able to find corrections to those results
obtained by applying the AdS/CFT correspondence to condensed matter physics.
It would be extremely interesting to explore this direction.

\section*{Acknowledgements}
This work was supported in part by the Grant-in-Aid for
Scientific Research Fund of the JSPS (C) No. 24540290, (C) No. 22540293,
and (A) No. 22244030.


\end{document}